\documentclass[twocolumn,showpacs,preprintnumbers,amsmath,amssymb]{revtex4}
\usepackage{graphicx}% Include figure files
\usepackage{dcolumn}% Align table columns on decimal point
\usepackage{bm}% bold math

\begin{document}

\preprint{MIFP-07}

\title{Supercriticality of a Class of Critical String Cosmological  Solutions}% Force line breaks with \\

\author{Dimitri V. Nanopoulos$^{1,2,3}$}
\author{Dan Xie$^1$}
\affiliation{$^{1}$George P. and Cynthia W.Mitchell Institute for
Fundamental Physics,\\ Texas A\&M University, College Station, TX
77843, USA \\ $^{2}$Astroparticle physics Group, Houston Advanced
Research
Center (HARC), Mitchell Campus, Woodlands, TX 77381, USA\\
$^{3}$Academy of Athens, Division of Nature Sciences,\\ 28
panepistimiou Avenue, Athens 10679, Greece
 }%

\date{\today}% It is always \today, today,
             %  but any date may be explicitly specified

\begin{abstract}

For a class of Friedmann-Robertson-Walker-type string solutions with
compact hyperbolic spatial slices formulated in the critical dimension,
we find the world sheet conformal field theory which involves the
linear dilaton and Wess-Zumino-Witten-type model with a compact
hyperbolic target space. By analyzing the infrared spectrum, we
conclude that the theory is actually supercritical due to 
modular invariance. Thus, taking into account
previous results, we conclude that all simple nontrivial string
cosmological solutions are in fact supercritical. In addition, we discuss the relationship
of this background with the Supercritical String Cosmology (SSC).

\end{abstract}

% PACS, the Physics and Astronomy
                             % Classification Scheme.
%\keywords{Suggested keywords}%Use showkeys class option if keyword
                              %display desired
\maketitle

\section{Introduction}
String theory is usually formulated in the critical dimension (D=26
for the bosonic string, and D=10 for the superstring ), and there
are various ways to justify why this is the natural choice\cite{Pol}. 
The idea of a critical dimension has been firmly entrenched in the
way that string theory is usually formulated, and it is widely
taken for granted that this should be the case in most studies.  
Nevertheless, it is possible to formulate string theory in a dimension other
than the critical dimension, and such a way of formulating string theory
finds a useful application in studying time-varying backgrounds such as those
which are necesssary for cosmology~\cite{emn1}\cite{emn2}.

There are several reasons to suspect that non-critical string theory
is relevant to the web of string duality. The first clue comes from
Cosmology. Observational data from the SnIa \cite{SI} project and
WMAP ${1,3}$ \cite{wmap1} strongly suggests that the 
universe is currently in an accelerating phase, which may be explained by the
presence of dark energy. String theory, as a theory of quantum
gravity, should give an explanation of the dark energy. However, it is very
difficult to find exact time-dependent solutions in  critical string
theory. However, it has been shown in \cite{ellis}\cite{bachas} that if one
goes beyond the critical dimension, one can find  simple
time-dependent Friedmann-Robertson-Walker (FRW) solutions with a
time-like linear dilaton. So-called supercritical strings (for strings with
dimension $D>D_{critical}$) may then also provide an explanation for the 
existence of
dark energy\cite{nano}.

From the theoretical point of view, past experience tell us that strings
probe space time in a very different way than we may intuitively think.
A compelling example is the emergence of M theory, where when we study the
strong coupling behavior of Type IIA string theory, the theory
develops a new dimension, and the theory is then formulated in $11$
dimensions! The extended nature of the string reveals many amazing
things about the property of space time (T duality is one of the
famous example). We expect that strings experience the space-time
dimensions in a novel way, and it may be possible that string theory
formulated in critical dimensions actually exhibits non-critical
behavior, i.e. the number of effective space time dimensions in
which strings can freely oscillate is larger than the critical
dimension.

In \cite{ellis}, the authors analyzed the conformal field
theory of FRW type solutions with space curvature $\kappa=0$ and
$\kappa=+1$; it was shown that the space dimensions should be
supercritical. In this paper, we will consider the case with space
curvature $\kappa=-1$. This solution did not attract much attention
because conformal field theory with hyperbolic target space is hard
to study. For this solution, we can choose the time-like linear
dilaton and the level of the hyperbolic CFT properly so that the
space time dimensions are critical. This solution is trivial if the
hyperbolic manifold is noncompact because the background is only
part of the flat Minkowski space. The amazing thing is that if we
make the hyperbolic manifold compact, (to make the solution
topologically nontrivial) the actual space time dimensions of this
background is supercritical. So we conclude for all the nontrivial
FRW type solutions, the theory should be formulated in non-critical
dimensions.

This paper is organized in the following way:  
In section II, we discuss the cosmological solutions derived from the
string equations of motion and the quantization of the supercritical
linear dilaton background. It is shown that for this simple time
dependent background, the infrared (IR) and ultraviolet (UV)
behaviors are changed drastically. In section III, we discuss the
stability of the cosmological solutions.  We find that the pseudotachyon modes do
not imply the instability of the background. In section IV, we study
the cosmological solutions with the compact hyperbolic spatial
section in critical dimension and we find that the theory is indeed
supercritical. In section V, the relation of this background with
supercritical string cosmology is pointed out.

\section{Cosmological Solutions and The Quantization}
To study the nontrivial space time background, it is quite useful to
start with the string world sheet non-linear sigma model in curved
space time. The bosonic string action reads:
\begin{eqnarray}
{\cal S}=-{1\over4\pi\alpha^{'}}\int_\sum d^2\sigma
\sqrt{-h}[h^{\alpha\beta}\partial_{\alpha} X_{\mu} \partial_{\beta}
X_{\nu}G^{\mu\nu}+\nonumber\\
 i\epsilon_{\alpha\beta}B^{\mu\nu}\partial_{\alpha}
X_{\mu}\partial_{\beta} X_{\nu}+ \alpha^{'}R\Phi(X)].
\end{eqnarray}
The equation of motion for the background fields can be derived from
the requirement that the $\beta$ functions of this two dimensional
field theory vanish to one-loop order, where the $\beta$ functions can be
found in \cite{beta}\cite{beta1}. We are interested in finding the
four dimensional FRW type solution (we assume that the four
dimensional CFT is decoupled from the internal Conformal Field
Theory (CFT)). We have found in \cite{ellis} the following simple
and $asymptotically$ $unique$ cosmological solutions (we only write
the four large dimension with the internal part of the background
suppressed) :

1. The Einstein static universe, with the
background fields as (the metric is expressed in the Einstein frame,
the relation of the metric between string frame and Einstein frame
is $g^E_{\mu\nu}=e^{-2\Phi/(D-2)}G^S_{\mu\nu}$)
\begin{equation}
{\Phi=\Phi_0,~~~~ b=2e^{-\Phi_0}\sqrt{\kappa}X^0}.
\end{equation}
Here we have a constant dilaton, while the axion $b$ is related to
the NS field strength through the duality relation
$H_{\lambda\mu\nu}
=e^{2\phi}\epsilon_{\lambda\mu\nu\rho}\bigtriangledown^\rho b$, and
the metric is given by
\begin{equation}
{(ds)^2=-(dX^0)^2 +[{dr^2\over{1-\kappa
r^2}}+r^2(d\theta^2+\sin^2\theta d\phi^2)]}.
\end{equation}

2. The second class of solutions involves both a nontrivial dilaton
and the non-trivial axion field. The solution is
\begin{equation}
 {\Phi=-2QX^0,~~~~ b=2Q^2\sqrt{\kappa}({e^{QX^0}\over Q})},
\end{equation}
while the metric is given by
\begin{equation}
{(ds)^2=-(dt)^2+t^2[{dr^2\over{1-\kappa
r^2}}+r^2(d\theta^2+\sin^2\theta d\phi^2)] }.
\end{equation}
The $\kappa$ is the curvature of the spatial slice, and $X^0$ is the
world sheet time coordinate, it is related to the Einstein time
coordinate as $t={1\over Q}e^{QX^0}$. The central charge deficit is
$\delta c=c_I-22\propto(1+\kappa)$, so only for $\kappa=-1$, can we
find the cosmological solutions in critical dimension, while for the
flat and positive curvature the theory is supercritical. However,
for $\kappa=-1$, the solution is trivial due to the fact that this
background is only part of the flat space. We can take a quotient of
this space to make the solution non-trivial; the remarkable thing is
that this background with the compact hyperbolic section is indeed
supercritical. Before we get into the discussion of the hyperbolic
case, it is very useful to give a review of the well-studied
solutions with $\kappa=0$ and $\kappa=1$, we will find some generic
features of time-dependent solutions.

\subsection{Spectrum of Cosmological Solutions with $\kappa=0$ and $\kappa=1$}
The simplest time dependent background is the so-called
supercritical linear dilaton background (SCLD), this solution can be
derived if we take $\kappa=0$ in our second class of solutions.  A discussion of this
background may also be found in\cite{meyr}. Regardless of its
simplicity, we can definitely learn a lot from studying this
background due to the nontrivial time dependence of the solution.
The background reads:
\begin{equation}
{G_{\mu\nu}=\eta_{\mu\nu}, ~~~~~~~~~\Phi=-2QX^0}.
\end{equation}
The central charge of this CFT is
\begin{equation}
{c=D-12Q^2}.
\end{equation}
The anomaly cancelation requires that $c=26$ for bosonic string, so
for $Q$ real, it is clear that the space time dimension is larger
than the critical dimension.

The effect of the linear dilaton is to change the world sheet
energy-momentum tensor to the form (where we have set $\alpha^{'}=2$)
\begin{equation}
{T_{zz}=-{1\over2}\partial_{z}
X^{\mu}\partial_{z}X_{\mu}-{Q}\partial_{z}^2X^0}.
\end{equation}
We can quantize the theory by using the conventional mode expansion, whic 
is possible, because the theory is Weyl invariant and
reparameterization invariant, and because we can gauge fix the
world-sheet metric as $h_{ij}=\delta_{ij})$.

The Virasoro generators are \cite{ellis}
\begin{equation}
{L_n=:{1\over2}\sum_ka^\mu_{n-k}a_{k\mu}:+i{Q}(n+1)a^0_n},
\end{equation}
where we consider the left
moving modes only; the right moving modes have the similar form.
The commutation relation for the creation and annihilation operators
are familiar:
\begin{equation}
{[a_m^\mu,a_n^\nu]=m\eta^{\mu\nu}\delta_{m+n}}.
\end{equation}
The hermiticity relations of the Virasoro generators $L_n^+=L_{-n}$
dictate that
\begin{equation}
{a^{\mu+}_n=a_{-n}^\mu +i2Q\eta^{\mu0}\delta_{n}}.
\end{equation}
This relation means that the zeroth component has a fixed imaginary
part:
\begin{equation}
{p^0=E+i{Q}}.
\end{equation}
The spectrum can be constructed by acting on the tachyon state with
the creation operators
\begin{equation}
{\mid\phi>=a_{-n_1}^{\alpha_1}......a_{-n_k}^{\alpha_k}\mid p;0>}.
\end{equation}
Using the method of OCQ quantization, we find the mass shell
condition to be
\begin{eqnarray}
m^2=-\vec{p}^2=-2+N+\bar{N}+i2Qp^0+\nonumber\\{Q}^2-i2EQ=
-2+N+\bar{N}-{Q^2}.
\end{eqnarray}

We see that the effect of the linear dilaton is to give a shift
$Q^2$ to the mass spectrum of the critical string. Now the spectrum
can be divided into three classes. Those states whose unshifted mass
lie in the range $m^2<0$ are the true tachyons. Those states whose
unshifted mass lies in the range $0<m^2<Q^2$ have negative mass
square now, we call them pseudotachyons as in \cite{eva2}. Those
states which have unperturbed mass square $m^2>Q^2$ are still
massive in the linear dilaton theory. Intuitively one might think
that the background is highly unstable due to the new emerging
negative mass squared modes. However, we will see in the following
section that those pseudotachyon modes are not so harmful as one
might think.

It is straightforward to extend the solution to the heterotic and
type II superstrings. The energy-momentum tensor and supercharge
become:
\begin{equation}
{T_B=-{1\over2}\partial X_{\mu}\partial X^\mu +Q\partial^2
X^0-{1\over2}\psi_\mu \partial\psi^\mu},
\end{equation}
\begin{equation}
{T_F=-\psi_{\mu}\partial X^\mu + 2Q\partial\psi^0}.
\end{equation}
The anomaly cancelation condition becomes
\begin{equation}
D-8Q^2=10.
\end{equation}
For the bosonic sector of the heterotic string, the anomaly
cancelation remains unchanged.

The quantization is similar to the bosonic case. One important
observation is that there is no mass shift to the fermionic modes.
This can be seen from several points of view. First the mass-shell
condition for a lowest-lying Ramond state is
\begin{equation}
 {(E^2+Q^2-\vec{p}^2)=-1+{(D-2)\over8}},
\end{equation}
Using the anomaly cancelation condition, we find $E^2-\vec{p}^2=0$.
Alternatively, this follows from the supercharge, whose moments are
\begin{equation}
{G_n=i\sum_k\psi^{\mu}_{n-k}X_{\mu,k}-2Q(k+{1\over2})\psi_n^0}.
\end{equation}
When acting on a highest weight state the zeroth moment becomes
\begin{equation}
{G_0=-i(\gamma_0E-\gamma\vec{p})},
\end{equation}
which is precisely the massless Dirac operator. From the
field-theory point of view, the quadratic piece of the lagrangian of
a fermion in the linear dilaton background is
\begin{equation}
{L=e^{2QX^0}(\bar{\psi}\partial_{\mu}\gamma^{\mu}\psi+m\bar{\psi}\psi)}.
\end{equation}
We need to rescale the fields so that it has the canonical kinetic
terms; the rescaled field $\tilde{\psi}=e^{QX^0}\psi$, however,
obeys the free Dirac equation in flat space-time without a mass
shift.

The quantization is similar for the $\kappa=1$ solution; the
corresponding world-sheet theory is a linear dilaton with a
Wess-Zumino-Witten (WZW) $SO(3)$ model, and the central charge of this
system is
\begin{equation}
c=1-12Q^2+{3k\over k+1}=4-12Q^2-{3\over k+1}.
\end{equation}
Thus, the theory is non-critical. A subtle point is that the WZW
curvature is different from the curvature in the Einstein frame: the
WZW curvature is given by $Q^2\kappa$ \cite{ellis}. A similar quantization of
this solution gives the same mass-shift to the whole spectrum
\cite{ellis}.

\subsection{Partition Function and Modular Invariance}

The partition function for the SCLD reads \cite{ellis}:
\begin{equation}
 {Z_D=V_D\int
 {d^Dk\over(2\pi)^D}(q\bar{q})^{-c/24-(-k^2+Q^2)/2}\sum_{oscillators}q^N\bar{q}^{\bar{N}}}.
\end{equation}
Setting $c=D-12Q^2$, the partition function can be easily calculated:
\begin{equation}
{Z(\tau)=iV_D(Z_X(\tau))^D,}
\end{equation}
where
\begin{equation}
{Z_X(\tau)=(8\pi^2\tau_2)^{-1/2}\mid\eta(\tau)\mid^{-2}},
\end{equation}
and $\eta(\tau)$ is the Dedekind $\eta$ function. Now, including the ghost part, the one-loop amplitude becomes:
\begin{equation}
 {Z_{T^2}=iV\int_F{d\tau d\bar{\tau}\over
 32\pi^2\tau_2^2}(8\pi^2\tau_2)^{(2-D)/2}(\eta\bar{\eta})^{2-D}}.
\end{equation}
Using the modular transformation for the $\eta$ function, it is easy
to show that the amplitude is modular invariant.

Next, we will analyze the IR and UV behaviors of the partition
function. The general form for the partition function looks like
\begin{equation}
{iV_D\int^\infty {d\tau_2\over2\tau_2}(8\pi^2\tau_2)^{-D/2}\sum_i\exp(-2\pi m_i^2\tau_2).}
\end{equation}
In the IR limit, which corresponds to $\tau_2\rightarrow \infty$, we
see that there is divergence due to the presence of a mode with
negative squared mass. For the  bosonic string, we have a tachyon in
the spectrum, which signals an instability of our theory, and we are
expanding around the wrong vacua. For the above cosmological
solutions, we have new pseudotachyon modes besides the original
tachyons. One may conclude that those modes signal the more severe
instability of our background. However, we will show below that this
is not the case due to the decreasing string coupling.

In the UV limit, since the dimensions that the string can oscillate in
are greater than that of the critical string, we expect the high
energy behavior to be modified, and the high energy density of
states of the theory to be changed. This is true as we can see from
direct calculations. The asymptotic density of states is (this can
be calculated by using the similar method in the critical string
\cite{bowi}\cite{aen})
\begin{equation}
{\rho(m)\rightarrow
e^{2\sqrt{2}\pi\sqrt{(D-2)/6}m}=e^{4\pi\sqrt{2+Q^2}m}}.
\end{equation}
The density of states for the critical string is reproduced when we
set $Q=0$.

Even if we did not know how to calculate the high energy density of
states directly, the supercritical behavior can be seen from the
modular invariance of the one-loop amplitude. The IR limit of the
partition function is controlled by the lowest lying negative mass
squared state with
mass-squared
$m^2=-2-Q^2$. The IR limit is
\begin{equation}
{Z_{IR}\rightarrow e^{2\pi\tau_2(2+Q^2)}},
\end{equation}
and due to the modular invariance under
$\tau_2\rightarrow{1\over\tau_2}$, the UV limit should be
\begin{equation}
{Z_{UV}\rightarrow e^{2\pi(2+Q^2){1\over\tau_2}}},
\end{equation}
which shows that the number of dimensions in which string can
oscillate is supercritical.

The above asymptotic form of the UV partition function can be
checked
by using the explicit form of the high energy
density of states $\rho(m)$ we get above:
\begin{equation}
  {Z_{UV}\rightarrow \sum_m\rho(m)e^{-2\pi\tau_2m^2}}.
\end{equation}
Summing over m by use of the saddle point approximation, we get that
\begin{equation}
{Z_{UV}\rightarrow e^{2\pi(2+Q^2){1\over\tau_2}}},
\end{equation}
which is exactly the same as we derived by using the modular
invariance.

Some comments are needed here. We have seen that the time dependence
of the background change the critical theory dramatically, in the IR
limit, there is a mass shift to the lowest mass modes; in the UV
limit, the high energy density of states is changed because the
string lives in the supercritical dimensions. There is no mystery
here. However, the relation between the IR behavior and UV
supercriticality  may provide a very novel way to generate new
dimensions.

The idea is that if we start from certain critical time dependent
backgrounds, we expect that there is mass shift to the critical
spectrum, i.e. the pseudotachyon modes appear in the spectrum. Then,
the background is actually supercritical from the requirement of
modular invariance of partition function.

Before finding an example, there are two questions that we should
answer. Since the pseudotachyons play a very important role in our
construction, the natural questions are, does our background with
the pseudotachyons make sense? Is it stable? This question will be
answered in the next section;  answers are positive and the
background is stable at least in the late time period!

Another question is that where the supercritical behavior comes from, 
or other words where are the hidden dimensions? This is essence of our
construction. Recall the famous example of T-duality, where we
compactify string theory on a circle. When we shrink the radius of
the circle, instead of losing a dimension, the winding modes will
restore the dimension and so in the dual theory we have the equivalent
theory with the same number of large dimensions. This provides a
clue to the source of the supercriticality: The winding modes will
provide the necessary degrees of freedom.

\section{Pseudotachyon and Its Resolution}

The first question we mentioned in last section was discussed first
in \cite{ellis} and~\cite{ellis1} (see also the discussion in
\cite{poly}, in which the pseudotachyon is called "good tachyon",
for recent discussion, see \cite{eva2}). We will give a short review
of the result. We have seen that in our time dependent
background, the time dependence will induce a mass shift to the
various string modes, so that there will be some negative mass-
squared particles if originally $ 0<m^2<{Q^2}$. Intuitively, these
modes may signal the instability of our background, but we will show
that these modes will not induce the large back-reaction to our
background. This can be seen first from the effective field theory
point of view. The quadratic piece of the lagrangian of a scalar
field in the linear dilaton background is
\begin{equation}
{L=e^{2QX^0}(-\eta^{\mu\nu}\partial_{\mu}\phi\partial_{\nu}\phi-m^2\phi^2)}.
\end{equation}
Then we rescale the field $\tilde{\phi}=e^{QX^0}\phi$ so that the
the kinetic term takes the canonical form, and $\tilde{\phi}$ obey
the free wave equation in the flat space and with a rescaled
physical mass $(m^2-Q^2)$. Then for the pseudotachyon with
$(m^2-Q^2)<0$, the modes $\tilde{\phi}$ is growing exponentially,
but the fields $\phi$ is coupled to the dilaton,
$\phi=e^{-QX^0}\tilde{\phi}$. We can then see that the exponentially
decreasing factor of the dilaton will compensate the growing of the
rescaled field $\tilde{\phi}$. So the pseudotachyons do not
necessarily signal classical instability.

The real problem with the tachyon is not the fact that they grow
exponentially with time, but that their back-reaction on the
originally configuration grows exponentially with time, so that
including this back-reaction moves us far away from our starting
point. However this is not the case for our pseudotachyons , because
our pseudotachyons couple to the other fields through the string
coupling $g_s=e^{<\phi>},$, and though the rescaled field grows
exponentially, taking into account of the string coupling, the back
reaction is in fact decays exponentially with time.

We can also consider this from the tachyon condensation point of
view. The one-loop divergence in the partition function of SCLD
shows that there is an instability toward producing
exponentially-growing pseudotachyon fields, as has shown in
\cite{eva2}, in the large $Q$ limit(this corresponds to the weakly
coupled region), the tachyon condensation process is in exact
agreement with the world-sheet RG flow, and since the corresponding
deformation operators for the RG flow are constructed from the
marginal and irrelevant operators. This fact means that the
condensation process will not change the central charge. Therefore,
the SCLD are consistent stable string backgrounds, even when the
loop corrections are taken into account.

\section{Supercriticality from Cosmological Solutions in Critical Dimension}
We have examined the solutions we outlined in the section II, and we
see that there is only one possibility where we can find a string
solution which is of the critical dimension, namely if we take
$\kappa=-1$, though we need to allow an imaginary B field. The
metric of this background is
\begin{equation}
{ds^2=-dt^2+t^2ds_{H_3}^2+ds^2_\perp},
\end{equation}
\begin{equation}
{\Phi=-2QX^0, b=i2Q^2({e^{QX^0}\over Q})},
\end{equation}
where $H_3$ is the $3$ dimensional non-compact hyperbolic space with
$\kappa=-1$, and $ds^2_\perp$ is a $6$ dimensional internal compact
space. The curvature of the WZW model is $-Q^2$ which is similar to
the positive curvature case. In reference \cite{eva1}\cite{eva2},
they consider a time dependent string frame background which is
rather different from our Einstein frame cosmological background
here; their backgrounds correspond to a accelerating expanding
version of our cosmological background in the Einstein frame.

Our background is a time dependent critical background with an
expanding spacial slices. Actually it is nothing but the inside of a
light cone in flat space $M_{1,3}$. We can formulate the world sheet
theory in analogy with the positive curvature case. The
corresponding world sheet sigma model is a non-compact
$SU(2,C)/SU(2)$ WZW model with the linear dilaton. This non-compact
WZW model has been studied in many papers in the context of AdS/CFT
correspondence \cite{seiberg1}, \cite{maldacena}, and for the
Euclidean case ( which is of our interest here), in
\cite{nwzw1}\cite{nwzw2}.

In the case of the SCLD background, the linear dilaton induces a
mass shift to the whole spectrum; we might expect that there also
are pseudotachyons for this background. However, there is a  mass
gap of the Laplacian on the hyperbolic manifold, so there are no
pseudotachyons in the noncompact theory. This is reasonable,
otherwise, the background is only part of the flat space. The linear
dilaton and the curved geometry is a description of flat Minkowski
background in a different gauge fixing procedure. Also if there are
pseudotachyon modes, the theory is inconsistent since there are no
extra central charge sources which can provide the necessary high
energy density of states.

This background is trivial; we take a subgroup $\Gamma$ (to make the
quotient space to be a compact manifold, there must be some
constraints on the $\Gamma$ \cite{hyper}) of the isometry group of
the Hyperbolic 3 manifold $PSL(2,C)$. The metric then is
\begin{equation}
  {ds^2=-dt^2+t^2ds^2_{M_3=H_3/\Gamma} + ds^2_{\perp}}.
\end{equation}
In the IR limit, the compactness of the manifold removes the gap of
the spectrum and we have the pseudotachyon; the similar analysis of
the modular invariance dictates that the actual number of dimensions
is supercrticial. The question is, what is the source for the extra
dimensions? Luckily, there is a theorem by Milnor \cite{milnor}
which states that the compact manifold of negative sectional
curvature has a fundamental group of exponential growth. Therefore
the winding modes can probably provide enough degrees of freedom
to make the theory supercritical. We will prove in the following
section that it is indeed the case.

\subsection{Hyperbolic manifold and the IR spectrum}
To simplify the discussion, we study the bosonic string in detail,
and the result can be easily generalized to the superstring. The metric
tensor of the hyperbolic manifold $H_n$ with $R=1$ takes the form:
\begin{equation}
 {ds^2=[dy^2+\sinh^2yd\Omega_{n-1}^2]}.
\end{equation}
The Laplace-Beltrami operator for this manifold reads:
\begin{equation}
 {\bigtriangleup={\partial^2\over\partial\sigma^2}+(n-1)\coth y{\partial\over\partial y} + (\sinh
 y)^2\bigtriangleup_{S^{N-1}}}.
\end{equation}
The eigenvalue equation is then $-\bigtriangleup\phi=\lambda\phi$,
The solution of this equation can be solved by separation of variables, and solutions can
be found in \cite{hyper} :
\begin{equation}
{\phi_\lambda=f_{\lambda}{(y)}Y_{lm}=\Gamma({n\over2})({\sinh
y\over2})^{1-n/2}P^{\mu}_{-1/2+ir}(\cosh y)Y_{lm}}.
\end{equation}
Where $Y_{lm}$ is the spherical harmonic on $S_{n-1}$ and
$P_\nu^\mu(x)$ is the associated Legendre functions of the first
kind. The constant $r$ is related to the eigenvalue through the
relation $ \rho_n=(n-1)/2, r=\sqrt{(\lambda-\rho^2_n)}$. The
asymptotic behavior for the radial wave function is
\begin{equation}
{f_{\lambda}(y)\simeq{2^n\Gamma(n/2)\Gamma(ir)\over
4\pi^{1/2}\Gamma(\rho_{n}+ir)}e^{-\rho_ny+iry}+c.c}.
\end{equation}
The radial functions will remain bounded at infinity provided the
parameter $r$ is real, so we find that there is a gap in the
spectrum i.e. $\lambda \geq\rho_n^2=(n-1)^2/4$, $\lambda\geq1$ for
$n=3$.

The coset $H_3^+=SL(2,C)/SU(2)$ is the set of all hermitian
two-by-two matrices $h$ with determinant one. We start with the
action \cite{nwzw1},
\begin{equation}
{S[hh^+]={k\over\pi}\int d^2z
(\partial_z\psi\partial_{\bar{z}}\psi+(\partial_z+\partial_z\psi)\bar{v}(\partial_{\bar{z}}+\partial_{\bar
 z}\psi)v)),
 }
\end{equation}
where $hh^+$ is the parametrization of the $H_3$:
\[
hh^+ = \left(
\begin{array}{cl}
  e^{\psi}(1+\mid v\mid^2)^{1/2} & ~~~v  \\
  \bar{v} & e^{-\psi}(1+\mid v\mid^2)^{1/2}

\end{array} \right).
\]
This is a conformal field theory with the central charge
$c_1=3k/(k-2)$, and the central charge of our system is then
\begin{equation}
c=1-12Q^2+{3k\over k-2}+c_I.
\end{equation}
This separation makes sense only when the hyperbolic-radius is
larger than the plank length, i.e. $k\gg1$. The level $k$ is related
to the Einstein frame space curvature as $1=1/(2Q^2k)$ \cite{ellis};
so the radius of the WZW target space is
$R={\sqrt{1/Q^2}}=\sqrt{2k}$.

To satisfy the anomaly cancelation condition, the dilaton can be
chosen as
\begin{equation}
{\Phi=-2QX^0,~~~~~~~~~~~Q=\sqrt{1\over{2(k-2)}}},
\end{equation}
Notice that the central charge of the linear dilaton is
$c_2=1-12Q^2$, and the central charge of the four dimensional
cosmological solutions is $c_1+c_2=4$, the theory is indeed
critical.

We now analyze the spectrum of the theory.  The mass-shell condition
of the this CFT reads:
\begin{equation}
{(L_0+\bar{L_0}-2)\mid \Psi,N,\tilde{N},p>=0},
\end{equation}
Where $L_0$ includes both the Euclidean $AdS$ part and the internal
part; for the $AdS$ part, the conformal weight of the states
\cite{maldacena1} is
\begin{equation}
{L_0\mid\psi>=-{j(j-1)\over k-2}|\psi>},
\end{equation}
where $j$ parametrizes the representations of the $SL(2,C)$, with
the value $j=1/2+is$, so the conformal weight of the states becomes
\begin{equation}
{\Delta(j)={1/4+s^2\over k-2}}.
\end{equation}
The minimum of the conformal weight is $\Delta(j)={1\over4(k-2)}$,
and similar analysis on the right hand modes give the same result.
The mass squared of this state is therefore
\begin{equation}
{m^2={1\over4(k-2)}+{1\over4(k-2)}-2-Q^2}.
\end{equation}
Since $Q^2={1\over2(k-2)}$, we see that the minimum mass squared of
the spectrum is $m^2=-2$, which is the same as that of the critical
bosonic string.

This result can also be seen from the harmonic analysis of $H_3^+$
space. The standard Sugawara construction gives the Hamiltonian
$L_0=-{1\over4(k-2)}\bigtriangleup$ where $\bigtriangleup$ denotes
the Laplace-Beltrami operator on $H_3^+$. We have proved that there
is a gap in the spectrum for the Laplace-Beltrami operator, so $L_0$
is bounded from below with the minimum value ${1\over4(k-2)}$.
Including the right hand modes and the linear dilaton, the result is
the same as we found above. We conclude that the minimum mass does
not change. This result is reasonable since our background is only
part of the flat space time.

Now we take a quotient of the $H_3$ and get a compact manifold. The
background is still the solution to the string equations of motion,
and our world-sheet theory involves the quotient of the previous
sigma model. It seems that the theory is still living in the
critical dimension, but careful analysis of IR and UV behavior
implies that our background is supercritical.

It will give us some insight about what is happening to our system
if we recall what the theory changes when we compactify sting theory
on a circle. One the one hand, the momentum of the center of motion
is quantized; on the other hand, the modular invariance of the
partition function requires the new contributions to the spectrum,
and those winding strings which wrap on the circle give the desired
excitations.

For the compact daughter space $M_3=H_3/\Gamma$, the situation is
rather similar: due to the compactness of the manifold, the spectrum
is discrete and the new novel feature is that the original gap has
been removed! The spectrum begins from $\lambda=0$
\cite{hyper4}(since there is no renormalizable conditions on the
wavefunction for the compact manifolds).

The quotient changes both the IR behavior and UV behavior of the
theory. In the IR limit, due to the disappearance of gap, the
minimum mass becomes $m^2=-2-{1\over2(k-2)}$; The mass-shift of the
minimum mass signals the emergence of new dimensions as we see from
the modular invariance condition; but where are the new dimensions?
The analogy with the circle case implies that the new dimensions may
hide as the winding strings. One clue indicates that it is indeed
the case. In the UV limit, due to the rich topology of the
hyperbolic manifold, i.e. the fundamental group is increasing
exponentially, the density of states of the theory is changed
drastically and the effect dimensions of the theory is indeed
supercritical. We will confirm this in the next section.

While we have only studied the bosonic string in the above analysis, the
same analysis can be easily applied to the bosons of the
superstring.  The conclusion is that the minimum mass-squared shifts
from zero to $m^2=-{1\over2(k-2)}$, and as the result of the
fermionic mode in SCLD background, there is no mass shift to the
fermions, so the minimal mass squared of the theory is
$m^2=-{1\over2(k-2)}$.

\subsection{Modular Invariance and Supercriticality}
Modular invariance is a very important consistency condition for
the string theory, and it also plays a important role in studying
the high energy density of states of the theory \cite{seiberg}.

Recall that, for our particular CFT, in the IR limit, the minimum
mass squared is
 \begin{equation}
 {m_{min}^2=-{1\over2(k-2)}}.
 \end{equation}
We are interested in the large $k$ limit where we can have the four
large macroscopic dimensions; The minimum is approximately
$m_{min}^2=-{1\over2k{\alpha^{'}}}$ (the $\alpha^{'}$ is included
from the dimensional analysis). The IR limit of the partition
function looks like
\begin{equation}
 {Z_{IR}\sim\exp{({\pi\tau_2\over
 2k})}},
 \end{equation}.
  Due to the modular invariance, the
  UV limit should look like
 \begin{equation}
 {Z_{UV}\sim\exp{({\pi\over2k\tau_2})}}.
 \end{equation}

From our study of the SCLD partition function, we can conclude that
the actual number of dimensions in which the string can freely
oscillate is supercritical.

We want to check this relation by direct calculation as we have done
in the SCLD case. There is a powerful tool that we can use to
calculate the asymptotical behavior of the partition function, the
Selberg trace formula (see \cite{hyper} for a detailed explanation
of the formula). The formula relates the quantum quantity to the
counting of the classical orbits. The Selberg trace formula is
\begin{eqnarray}
\sum_{j=0}^{\infty}h(r_j)=\Omega(F_n)\int_0^{\infty}h(r)\Phi_n(r)dr+\nonumber\\
\sum_{\{\gamma\}}\sum_{n=1}^{\infty}{\chi^{n}(\gamma)l_{\gamma}\over
S_3(n;l_\gamma)}\hat{h}
  (nl_\gamma),
\end{eqnarray}
where $h$ is a function defined on the compact hyperbolic manifold
with some analytic constraints; $l_\gamma$ is the length of the
nontrivial cycle. $S_3(n;l_\gamma)$ is a definite function of $n$
and $\l_\gamma$. The summation over $n$ is a sum over winding
numbers and summation over $\gamma$ is a sum over all the nontrivial
cycles.

We can use this trace formula to compute $Tr\exp(\Delta_{\Gamma})$
over the $L^2(H_3/\Gamma)$, which is essential in the calculation of
the partition function. Then
$\hat{h}(nl_\gamma)=e^{-(nl_\gamma)^2\tau_2/4\pi\alpha^{'}}$. In the
UV limit, the summation over $n$ and $\gamma$ can be transformed to
a integral over the length of nontrivial circle. In this limit, the
function $S_3(n,l_\gamma)$ is simply
\begin{equation}
{1\over S_3(n,l_\gamma)}=({1\over\sinh(l/2\sqrt{2k\alpha^{'}})})^2,
\end{equation}
where we have used the fact that $R=\sqrt{2k\alpha^{'}}$; and now
$\hat{h}(nl_\gamma)=e^{-\tau_2l^2/4\pi\alpha^{'}}$. According to the
Milnor's theorem, the density of periodic geodesics is given by
\cite{hyper1}\cite{hyper2}\cite{hyper3}:
\begin{equation}
{\rho(l)={1\over l}e^{2l/\sqrt{2k\alpha^{'}}}}.
\end{equation}
Altogether, the UV partition function is then
\begin{equation}
{\int dl
e^{2l/\sqrt{2k\alpha^{'}}}({1\over\sinh(l/2\sqrt{2k\alpha^{'}})})^{2}e^{-l^2\tau_2/(4\pi
\alpha^{'})}}.
\end{equation}

Using the saddle point approximation, this integral gives the result
\begin{equation}
Z_{UV}\sim \exp({\pi\over 2k\tau_2}).
\end{equation}.
This is in exact agreement with the result from the modular
invariance.

So far we only consider the bosonic sector. For the fermions, as we
have seen in the Heterotic and Type II case, there is no mass shift
to the fermionic modes due to the linear dilaton. So we would expect
that there is no exponential contributions in the UV limit, so that
there is no boson fermion cancelation (See also the discussion in
\cite{eva1} ). It is interesting to find a specific co-compact
subgroup of $PSL(2,C)$ and calculate the exact partition function.

We have shown that for this specific critical background with
compact negative curvature slice the theory is indeed supercritical.
One may wonder whether we can find a dual description in which the
supercriticality can be manifest in a geometrical description like
we usually see in the linear dilaton theory. We may find a dual
conformal field theory in which the supercriticality is manifest by
using different kinds of gauge fixing procedure along the line
\cite{horova}. For the accelerating expanding background
\cite{eva1}, there is a paper \cite{eva3} which appeared recently, which
shows that the "D-dual" description of a Riemann surface of genus
$h$ can be expressed in terms of its $2h$ dimensional Jacobian
torus, perturbed by a closed string tachyon arising as a potential
energy term in the worldsheet sigma model (see also the discussion
on the dualized D brane in \cite{dbrane}). Since our manifold
contains a $AdS$ factor (the Euclidean version), we may study the
dual theory from the $AdS/CFT$ correspondence as discussed also in
\cite{eva3}.

Another interesting observation of the accelerating expanding
solution is that we have the dynamical dimension reduction. Since
the central charge is decreasing with time (the dilaton is
$\phi=-2QX^0$, and $Q(t)={1\over t}$ \cite{eva1}, here $t$ is the
Einstein time and it is rescaled so that $t=R$, where $R$ is the
radius of the hyperbolic space; note that for our background,
$Q=\sqrt{1\over2(k-2)}={1\over R}$ in the large k limit), and from
the dual picture, the geometric dimensions are decreasing due to the
tachyon condensation. For detailed description of dimension changing
solutions in the supercritical string theory, see
\cite{hellerman},\cite{hellerman1}. One of important questions about
string cosmology is to understand why our universe can reduce down
to four dimensions while the string theory is usually formulated in
ten dimensions. The solution discussed above demonstrates that the
cosmological solution can dynamically reduce the space dimension.
This may give a dynamical explanation of why we are living in four
dimensions. A natural question is why the transition stops at $D=4$.
A possible explanation is provided in \cite{aen}. As noted in
section II, the high energy density of states is
\begin{equation}
\rho(m)\propto m^{-D}\exp^{4\pi\sqrt{Q^2+2}m}.
\end{equation}
A level density of this type leads to a phase transition at a
temperature $T={1\over4\pi\sqrt{Q^2+2}} $\cite{hage}. As the
universe is cooling down, there are a couple of phase transitions
which dynamically reduce the space time dimensions. The dynamical
dimension reduction must stop at $D=4$ as shown in \cite{aen},
because for this level density, the pressure and density is infinite
for $D<4$ near the transition temperature, while the pressure and
density for $D\geq4$ is finite near the transition point, so the
state with $D<4$ can not be obtained from a phase with finite
pressure and density i.e. $D=4$. The universe stops at $D=4$.

\section{The relation with the Supercritical string cosmology}
If the universe is described by critical string theory, the total
dimensions of the flat spatial coordinates and the internal space
must add up to $c_{tot}=25$($\hat{c}_{tot}=9$ in superstring) at the
present stage of evolution. As we mentioned as in the Introduction,
it is very hard to find a critical solution describing the
accelerating universe.

However, one may speculate that the total number of space-time
dimensions has been larger or smaller than $25$ (for clarity, we
take the bosonic string as the example) in the early universe,
\cite{ellis}. Moreover, one may propose that the present universe is
in a supercritical phase \cite{brany}\cite{dimitri}.  With this
assumption, we can find the accelerating solutions. This proposal is
supported in \cite{tyestin}, which proves that the space of general
time-dependent solutions of classical bosonic string theory contains
attractors which are static solutions with spatial CFT parts which
are minima of $\mid\bar{c}-25\mid$, where $\bar{c}$ is a generalized
Zamolodchikoc's c-function.

Our particular model supports this proposal, and corresponds to the
case $c_{tot}=25$ {or $\hat{c}=9$ for the superstring}, where the critical
limit ($t\longrightarrow\infty$, flat space limit) is the fixed
point of our time-dependent solution, and at the present time the
universe is in a supercritical phase with central charge $c>25$.

For the general case with fixed point $c=25$, the equations at the
vicinity of this fixed point are
\cite{emn1}\cite{emn2}\cite{tyestin}\cite{dimitri}:
\begin{equation}
\ddot{\vec{\lambda}}+Q(t)\dot{{\vec{\lambda}}}+O(\dot{\lambda}^2)=-\vec{\beta},
\end{equation}
where $\vec{\lambda}$ is the background fields and $\beta$ is the
world-sheet $\beta$ function, i.e. the time evolution of the
solutions is the same as the world sheet renormalization group flow.
It has been shown that for the accelerating expanding model
$Q(t)=1/t$ \cite{eva1}, and one interesting thing is that in
\cite{brany}\cite{dimitri}\cite{tyestin} it has been shown that for
the $SO(3)$ WZW model with $c_{tot}=25$, $Q(t)$ is proportional to
$1/t$.

\section{Discussion and Conclusion}
In this paper, we find that the simple non-trivial cosmological
solutions with the compact hyperbolic spatial slices, although
formulated in critical dimension, are indeed supercritical. With the
previous results on the $\kappa=0$ and $\kappa=1$ case, we may
conclude that for all the simple non-trivial cosmological solutions,
the theory is indeed supercritical. In the framework of the
non-critical string theory, the space time dimensions can be reduced
dynamically through the tachyon condensation, and we found a
possible explanation of why we are living in $D=4$: the transition
must stop at $D=4$ because the temperature and pressure is infinite
near the transition point for $D<3$ while those thermodynamical
quantities are finite for $D\geq4$.  A state with infinite
temperature and pressure cannot be obtained from a state with finite
thermodynamical quantities.

We also found an interesting way to generate the supercritical
behavior from the critical string, namely to produce the
pseudotachyon. We analyzed an example that exhibits the desired
behavior. The simplicity and solidity of our treatment suggests that
supercritical behavior is very important in studying the time
dependent background in string theory. It is interesting to see
whether the similar phenomenon appears in the conventional
compactification like flux compactification.

The above phenomenon is very interesting, from the duality point of
view. We have found that degrees of freedom coming from the winding
modes give new contributions to the effective central charge, and
the theory is indeed supercritical. There may exist a dual theory in
which the geometric dimensions is different from the original
theory. The main difference of this duality is, the two dual
theories may have different space time dimensions. This is a new type of
duality: D-duality \cite{eva3}.

Some generalization are in order.  First, it is not necessary that
the manifold must be compact; as long as there is a compact section in
the manifold, it is possible to have the supercritical behavior.
Second is the question of whether we can derive the conventional duality like the heterotic
$K3$ duality from this mechanism. If we can think of $K3$ surface as
a manifold containing a hyperbolic section and we take a suitable
dilaton profile, we may give a semi-classical explanation of the
$K3$ duality.

This model may also provide a possible solution to supersymmetry breaking in
string theory. In \cite{ellis}\cite{pol}, it is proposed that
the charge deficit coming from the internal manifold can generate a
tree level cosmological constant, but the problem is that to get a
small cosmological constant since the internal manifold cannot be kept at
the small size, the internal dimensions will decompactify. Now if we
take the compact hyperbolic manifold $H^3/\Gamma$ as the spatial
slice of our Universe. Since the theory is supercritical and we have
the potential term coming from the winding string, supersymmetry
is broken (this mechanism may signal the high energy scale SUSY
breaking). For this kind of model we do not have the
decompactification problem. See also \cite{brany}\cite{dimitri}.

From the cosmological point of view, since the central charge of our
system is changing with the time, to restore the full conformal
invariance, the dilaton is sourced, we have the time varying central
charge deficit $Q^2(t)$. This is just the description of Q cosmology
introduced in the reference \cite{dimitri}. One may worry that the
space manifold must be compact so that we have the supercritical
behavior. This is not the necessary condition, localized
psuedotachyon can produce the supercritical behavior as well, and
those psuedotachyons can possibly be produced in the early universe,
for instance, in the inflationary scenario. In string theory, those
modes can be produced from the catastrophic events like big bang,
brane collisions, etc.

\begin{acknowledgments}
We would like to thank Eric Mayes for a critical reading of the manuscript.  
This research was supported in part by the Mitchell-Heep Chair in High Energy Physics,
and by the DOE grant DE-FG$03$-$95$-Er-$40917$(DVN).

\end{acknowledgments}

\end{document}